\documentclass[final,3p,times,twocolumn]{elsarticle}

\usepackage{natbib}

\usepackage{graphicx}

\usepackage{amssymb}

\journal{Nuclear Instruments and Methods A}

\begin{document}

\begin{frontmatter}


\title{Performance of a MICROMEGAS-based TPC in a high-energy neutron beam}

\author[LLNL]{L. Snyder\corref{corresponding}}
\cortext[corresponding]{Corresponding author}
\ead{snyder35@llnl.gov}

\author[LANL]{B.~Manning}
\author[LLNL]{N.~S.~Bowden}
\author[CSM]{J.~Bundgaard}
\author[LLNL]{R.~J.~Casperson}
\author[UC]{D.~A.~Cebra}
\author[LLNL]{T.~Classen}
\author[LANL]{D.~L.~Duke}
\author[UC]{J.~Gearhart}
\author[CSM]{U.~Greife}
\author[LLNL]{C.~Hagmann}
\author[LLNL]{M.~Heffner}
\author[CSM]{D.~Hensle}
\author[CSM]{D.~Higgins}
\author[ACU]{D.~Isenhower}
\author[OSU]{J.~King}
\author[CAL]{J.~L.~Klay}
\author[LANL]{V.~Geppert-Kleinrath}
\author[OSU]{W.~Loveland}
\author[LLNL]{J.~A.~Magee}
\author[LLNL]{M.~P.~Mendenhall}
\author[LLNL]{S.~Sangiorgio}
\author[LLNL]{B.~Seilhan}
\author[LANL]{K.~T.~Schmitt}
\author[LANL]{F.~Tovesson}
\author[ACU]{R.~S.~Towell}
\author[ACU]{S.~Watson}
\author[OSU]{L.~Yao}
\author[LLNL]{W.~Younes}

\author[]{\protect\\(The NIFFTE Collaboration)}

\address[LLNL]{Lawrence Livermore National Laboratory, Livermore, CA 94550, United States}
\address[LANL]{Los Alamos National Laboratory, Los Alamos, NM 87545, United States}
\address[CSM]{Colorado School of Mines, Golden, CO 80401, United States}
\address[UC]{University of California, Davis, CA 95616, United States}
\address[ACU]{Abilene Christian University, Abilene, TX 79699, United States}
\address[CAL]{California Polytechnic State University, San Luis Obispo, CA 93407, United States}
\address[OSU]{Oregon State University, Corvallis, OR 97331, United States}


\begin{abstract}

The MICROMEGAS (MICRO-MEsh GAseous Structure) charge amplification structure has found wide use in many detection applications, especially as a gain stage for the charge readout of Time Projection Chambers (TPCs). Here we report on the behavior of a MICROMEGAS TPC when operated in a high-energy (up to $800~$MeV) neutron beam. It is found that neutron-induced reactions can cause discharges in some drift gas mixtures that are stable in the absence of the neutron beam. The discharges result from recoil ions close to the MICROMEGAS that deposit high specific ionization density and have a limited diffusion time. For a binary drift gas, increasing the percentage of the molecular component (quench gas) relative to the noble component and operating at lower pressures generally improves stability.
\end{abstract}


\end{frontmatter}


\section{Introduction}
A MICROMEGAS (MICRO-MEsh Gaseous Structure) provides high-gain, high-speed signals with good spatial resolution from a structure with low areal density~\cite{Giomataris}.  The MICROMEGAS has therefore been widely used in gaseous detectors, especially Time Projection Chambers (TPCs)~\cite{Nygren}. In this latter case, the drift gas must be suitable for both the amplification stage and the TPC tracking application, i.e. provide sufficient stopping power, low diffusion and low recombination. 

The NIFFTE (Neutron-Induced Fission Fragment Tracking Experiment) collaboration has constructed the fissionTPC, a compact dual volume MICROMEGAS TPC, to perform precision neutron-induced fission cross section measurements~\cite{Heffner}. Unusually, the fissionTPC will be operated in a high-energy neutron beam. While non-tracking MICROMEGAS-based detectors have been previously developed as neutron detectors and to make beam profile measurements~\cite{Belloni,Pancin,Andriamonje}, there are no dedicated studies of discharging in  MICROMEGAS-based TPCs that are operated directly in a high-energy neutron beam. The MICROMEGAS based $\mu$TPC~\cite{Golabek} was operated at energies below $1$~MeV, almost 3~orders of magnitude lower energy than the maximum energy generated by the LANSCE (Los Alamos Neutron Science CEnter) facility at which the fissionTPC is operated. 

Having a neutron beam passing directly through the MICROMEGAS is a cause for concern, since charged particle tracks resulting from interactions such as (n,n$^\prime$) and (n,p) can originate in or very near the gain stage.  The close proximity of the track to the MICROMEGAS limits the diffusion time and can result in high density electron clouds that cause breakdown via the mechanism described by Raether~\cite{Raether}.  Therefore, we have performed a stability study of the fissionTPC MICROMEGAS with several common drift gases operated at gains and pressures relevant to the detection of charged particles resulting from neutron-induced reactions.  The gas mixtures used were Ar based with C$_4$H$_{10}$, CH$_4$, CF$_4$, and CO$_2$ as quench gases. 

In Section~2 we describe the experimental conditions, including the LANSCE neutron beam facility, the fissionTPC MICROMEGAS, the gas handling system of the fissionTPC, and the experimental procedures used. Results from the gases used are reported in Section~3. Finally, in Section~4 we provide discussion and interpretation of the results, describing a possible mechanism for gain instability and identifying characteristics of gas mixtures that are suitable for use in high-energy neutron beam experiments.  

\section{Experimental Setup}

\subsection{Neutron Beam}
The Weapons Neutron Research Facility (WNR) at LANSCE was used for these measurements~\cite{Lisowski}. 
Spallation neutrons are produced by impinging a $\sim$5~$\mu$A beam of 800~MeV protons onto a cylindrical tungsten target ($7$~cm long with a $3$~cm radius). 
The neutrons are only moderated by a thin water jacket surrounding the tungsten target. 
The proton beam at WNR is pulsed, resulting in neutron macropulses $625$~$\mu$s long delivered at $100$~Hz.  
Within each macropulse, there are micropulses seperated by $1.8~\mu$s.
The 90L flight path at WNR was used, a plot of the neutron flux is shown in Fig. 2 of Ref.~\cite{Tovesson}.

\subsection{MICROMEGAS Design}
\begin{figure}[ht]
\centering\includegraphics[width=1.\linewidth]{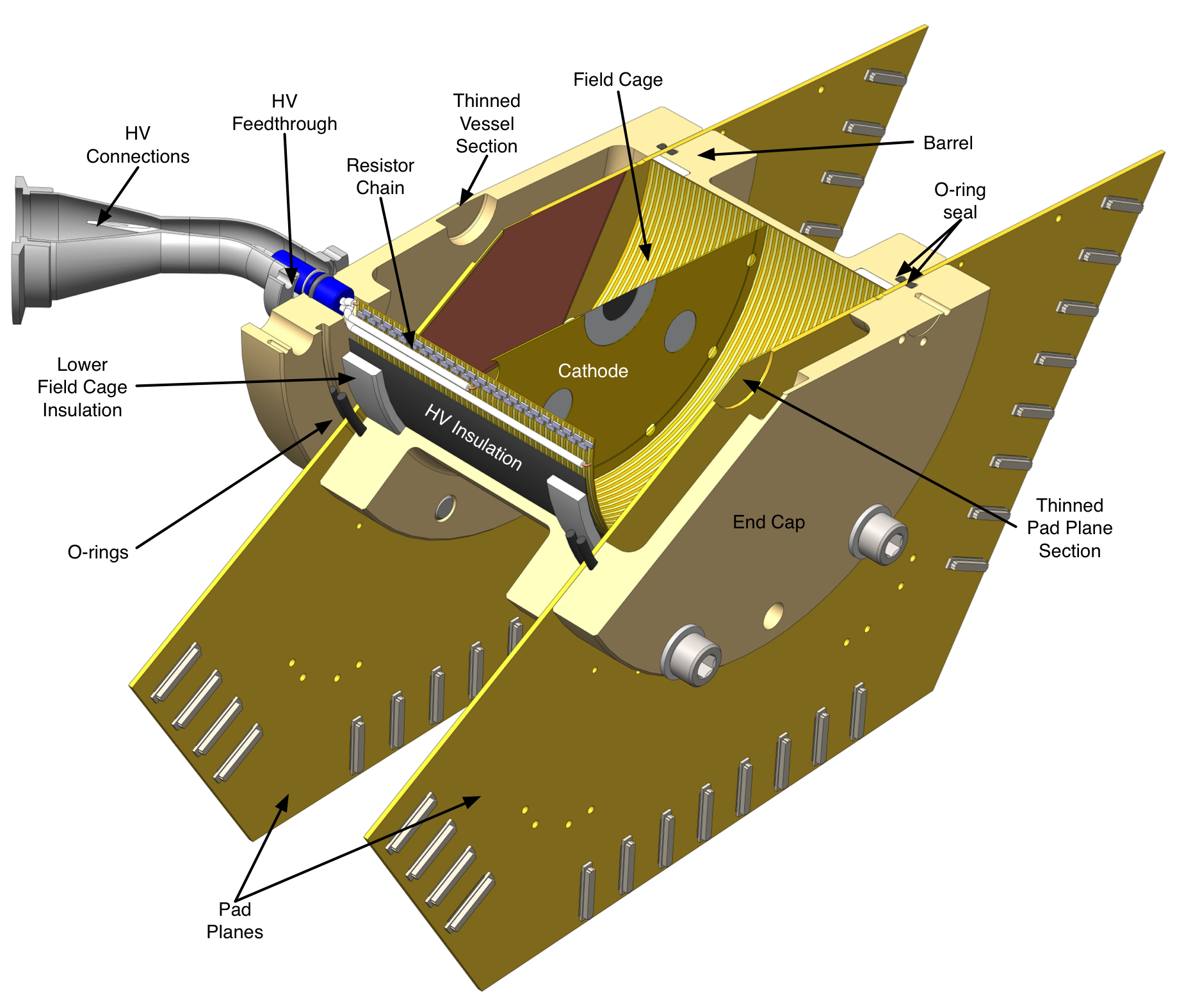}
\caption{\label{fig:cutaway} A cutaway image of the fissionTPC.  The neutron beam passes through the thinned sections of the vessel and pad plane.  The actinide target is mounted in the center of the cathode.  This figure was taken from Fig.~2 of Ref~\cite{Heffner}.}    
\end{figure}

\begin{figure}[ht]
\centering\includegraphics[width=1.\linewidth]{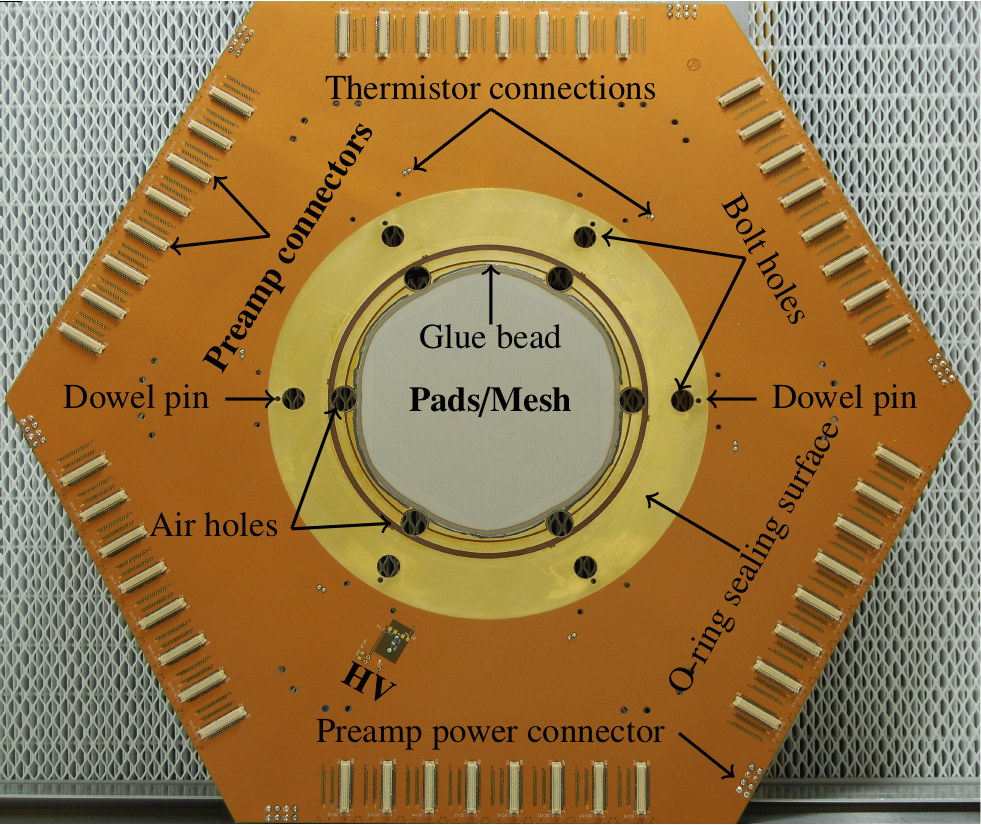}
\caption{\label{fig:padplane} The fissionTPC pad plane.  The MICROMEGAS mesh covers the pads at the center.  This figure was taken from Fig.~5 of Ref~\cite{Heffner}.}
\end{figure}

The fissionTPC has two volumes separated by a central cathode~\cite{Heffner}.  
The total gas volume is  $\sim2$~liters with a 5.4~cm drift distance between each pad plane and the central cathode.  
The pad planes are 16-layer printed circuit boards with a total of 5952 hexagonal pads, each 2~mm in pitch.  
The MICROMEGAS consists of an electroformed nickel mesh, 3~$\mu$m thick with 1000 lines per inch and a diameter of 140~mm.  
The mesh is held above a pad plane surface by dry film solder mask pillars 75~$\mu$m tall and 500~$\mu$m in diameter, and is secured to the perimeter with a conductive epoxy. 
The electric field in the drift and gain regions can be controlled independently. 
A cutaway of the fissionTPC is shown in Fig.~\ref{fig:cutaway} and a picture of the pad plane with a mesh attached can be seen in Fig.~\ref{fig:padplane}

\begin{figure}[ht]
\centering\includegraphics[width=1.\linewidth]{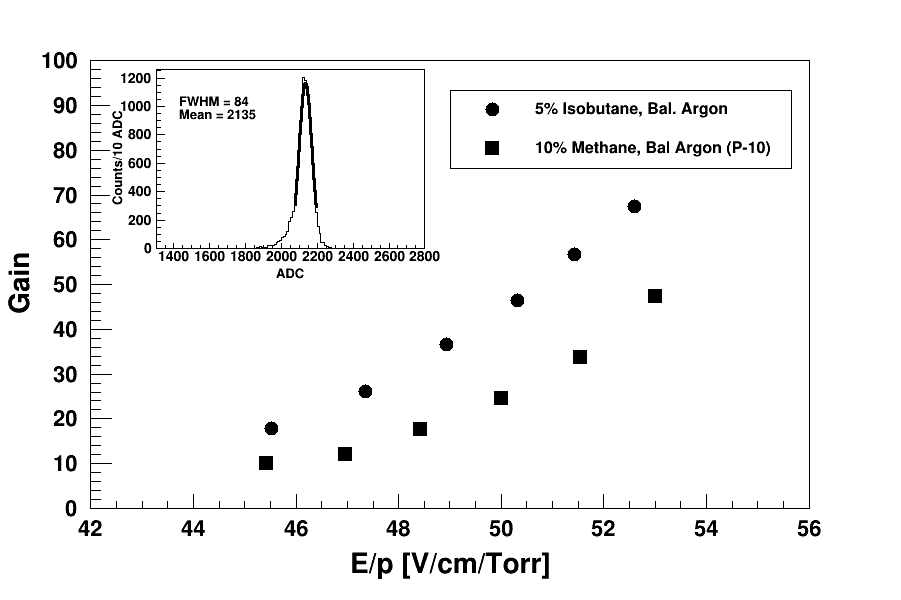}
\caption{\label{fig:gain} The MICROMEGAS gain as a function of the reduced electric field in the fissionTPC for Ar with 5\% C$_4$H$_{10}$ (circles) and 10\% CH$_4$ (squares).  A example fit of the $\alpha$ peak of a $^{252}$Cf source is shown in the inset. The fit was used to determine the gain and energy resolution.  This figure was taken from Fig.~9 of Ref.~\cite{Heffner}.}
\end{figure}

At the typical operating pressures of the fissionTPC (500--900~Torr), fission fragments will be stopped within several centimeters while depositing their full energy into the gas.  The total kinetic energy released in neutron-induced fission is $\sim~170$~MeV~\cite{Meierbachtol,Yanez}, with that energy being shared between two fragments. The short track lengths and high kinetic energies of fission fragments results in high-density primary electron clouds (about 10$^6$ electron--ion pairs) which allows the fissionTPC MICROMEGAS to operate at relatively low gains, ranging from 10--50.  Fig.~\ref{fig:gain} shows a gain curve for two typical operating gases as a function of ratio of the applied electric field, $E$, to the gas pressure, $p$ (the reduced electric field).

\subsection{Gas System}
The fissionTPC has a once-through gas system. The gas flow rates and pressure are maintained by an MKS~647C controller with an internal PID controller~\cite{MKS}.  The 647C adjusts the flow rates based on a user-defined pressure setpoint, while the mixture is  maintained by a flow ratio set by two inlet Mass Flow Controllers (MFC), one for  Ar and one for the quench gas.  The exchange rate is set by an independently controlled MFC on the exhaust side.  The MFCs were MKS~1479A, and the pressure transducer used for the PID feedback signal was an MKS~627B.  The Ar and quench gas inlet MFCs have a full scale of 1000~SCCM and 200~SCCM respectively, with an accuracy of 1\% of full scale.

A calculation of the mixing rate shows that for an exchange rate of 300~SCCM the 2~L gas volume will achieve 99.99\% of the desired composition in approximately 60 minutes. For each gas mixture in the experiment, the gas was exchanged for at least 1 hour at a rate of 300~SCCM before testing commenced.  The purity of the Ar and CF$_4$ gas was 99.999\%, the CO$_2$ was 99.99\%, and the C$_4$H$_{10}$ was 99\% with $<$ 5 ppm of water and the remaining contaminants consisted primarily of other alkanes.

\subsection{Experimental Procedures} \label{procedures}
At the time of this experiment the fissionTPC was configured to measure neutron-induced fission cross sections and contained a target with deposits of $^{239}$Pu and $^{235}$U on the cathode.  The $\alpha$ activity of the $^{239}$Pu was approximately 0.425~MBq, with a comparatively negligible activity from the $^{235}$U of 10 Bq.   The measurements, unless otherwise noted, took place at a pressure of 760~Torr.  This pressure was chosen such that for any gas, ranging from pure Ar to pure quench gas, the $^{239}$Pu $\alpha$-particle tracks would be fully contained within the active area of the detector.  The signal size of the peak corresponding to full energy deposition of $^{239}$Pu $\alpha$-particles was then used to determine the gain of the MICROMEGAS.  The gain was adjusted to be the same for each gas mixture within 20\% and was set near typical operating conditions for the fissionTPC for fission cross-section measurements.  Additional data were collected at different pressures and higher gains to investigate the limits of MICROMEGAS stability.

Once the gain was set, the neutron beam shutter was opened and the MICROMEGAS was monitored for sparking, which is observed by a current trip of the high voltage system.  The trip limit was set to 500~nA.  The system was considered stable if no trips were observed after 60 minutes of exposure to the neutron beam.  If a trip was observed in less than 60 minutes, the test was repeated for confirmation and a shutter-closed test of 60 minutes was conducted to confirm the instability was beam induced.
The experiment is thus a measure of a the threshold above which sparking will begin to occur rather than a measure of a sparking rate or probability.

\section{Results}
The primary measurements for comparison between different gas types were taken at $p=760$~Torr, with the gain for each gas mixture adjusted to be similar (Table~\ref{results_table}).  Stability against sparking in the beam is primarily a function of the quantity of quench gas in the mixture.  C$_4$H$_{10}$ and CO$_2$ were stable with concentrations at or above 1\%, CF$_4$ required a concentration of 25\% or greater, and CH$_4$ was only tested at a concentration of 10\% and found to be unstable.  The following sections further describe the data collected at various gains and pressures.

\begin{table}[htb]
\begin{center}
\begin{tabular}{cccc}
\hline
\rule{0pt}{3ex}%
\vspace{0.1cm}
\textbf{Quench Gas} & \textbf{Stability Conc.} & $\mathbf{E/p}$ (V/cm/Torr) & \textbf{Gain}\\
\hline
\rule{0pt}{3ex}%
C$_4$H$_{10}$ & 1\% & 33.3 & 21.8\\
\rule{0pt}{3ex}%
CO$_2$ & 1\% & 49.7 & 17.1\\
\rule{0pt}{3ex}%
CF$_4$ & 25\% & 61.2 & 18.8\\
\rule{0pt}{3ex}%
CH$_4$ (P10) & -- & 50.7 & 21.2\\
\end{tabular}
\end{center}
\caption{Stability limits for mixtures of Ar with C$_4$H$_{10}$, CO$_2$, CF$_4$, CH$_4$.  The stability concentration is the percent of quench gas at or above which the MICROMEGAS was stable against sparking in beam.  These data were collected at $p=760$~Torr and the given reduced electric field and corresponding gain values. }\label{results_table}
\end{table}

\subsection{CH$_{4}$} \label{p10section}
The initial gas explored for use in the fissionTPC was P10, a mixture of 90\% Ar and 10\% CH$_4$.  P10 is relatively cheap, provides fast drift speeds, and is commonly used in ionization chambers.  When operating with P10 out of beam, the fissionTPC was found to be highly stable, running for hundreds of hours over a wide range of gains and pressures.  P10 was found to be unstable to sparking in beam at $p=760$~Torr with a gain of $\sim$20.  A stable point was found when operating at a lower pressure of 530~Torr.  Table \ref{methane_table} shows the results of two tests at $p=530$~Torr, where the reduced electric field was varied to identify the transition to instability. At this lower pressure the gain could not be directly measured because the $\alpha$-particle tracks extended beyond the active area of the detector.  

\begin{figure}[ht]
\centering\includegraphics[width=1.\linewidth]{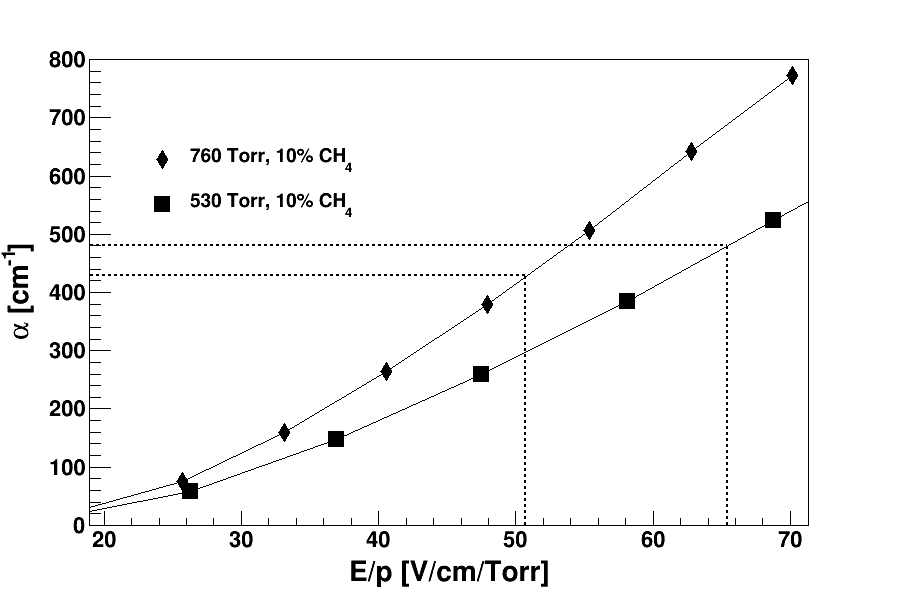}
\caption{\label{fig:gMeth} A Magboltz simulation of the first Townsend coefficient, $\alpha$, as a function of the reduced field, $E/p$, for two pressures of an Ar--10\% CH$_4$ gas mixture.  The stable configuration of $p=530$~Torr and $E/p=65.4$ is at a higher $\alpha$ (and therefore gain) than the unstable configuration of $p=760~$Torr and $E/p=50.7$ (dashed lines).}
\end{figure}

A Magboltz simulation~\cite{Biagi} was completed to estimate the dependence of $\alpha$, the first Townsend coefficient, on the reduced electric field. This is shown for the two gas pressures tested in Fig.~\ref{fig:gMeth}. Note that $\alpha/p$ scales with $E/p$~\cite{Aoyama,Sharma} and gain scales exponentially with $\alpha$.  
One can see that for $p=530$~Torr the stability point, $E/p=$~65.4, is at a higher $\alpha$ value (and therefore gain) than the point at $p=760$~Torr and $E/p=~50.7$ that was found to be unstable.  This indicates that sparking is not only a function of gain but also ionization density.  At lower pressures the stopping power is decreased, resulting in longer particle track lengths and a reduced primary ionization density.  Therefore higher gain can be applied before the Raether limit is reached.

In addition to being stable out of beam, the system was stable in beam, with the drift field set to 0~V/cm and a nominal field in the gain stage. 
The interpretation of this result will be discussed further in Section~\ref{discussion}.

\begin{table}[htb]
\begin{center}
\begin{tabular}{cccc}
\hline
\rule{0pt}{3ex}%
\vspace{0.1cm}
\textbf{p (Torr)} & \textbf{Conc.} & $\mathbf{E/p}$ (V/cm/Torr) & \textbf{Stable}\\
\hline
\rule{0pt}{3ex}%
530 & 10\% & 65.4 & Yes\\
\rule{0pt}{3ex}%
530 & 10\% & 67.9 & No\\
\rule{0pt}{3ex}%
760 & 10\% & 50.7 & No\\
\end{tabular}
\end{center}
\caption{Configurations and results for stability against sparking for an Ar--CH$_{4}$ mixture.}\label{methane_table}
\end{table}

\subsection{C$_4$H$_{10}$}
\begin{figure}[ht]
\centering\includegraphics[width=1.\linewidth]{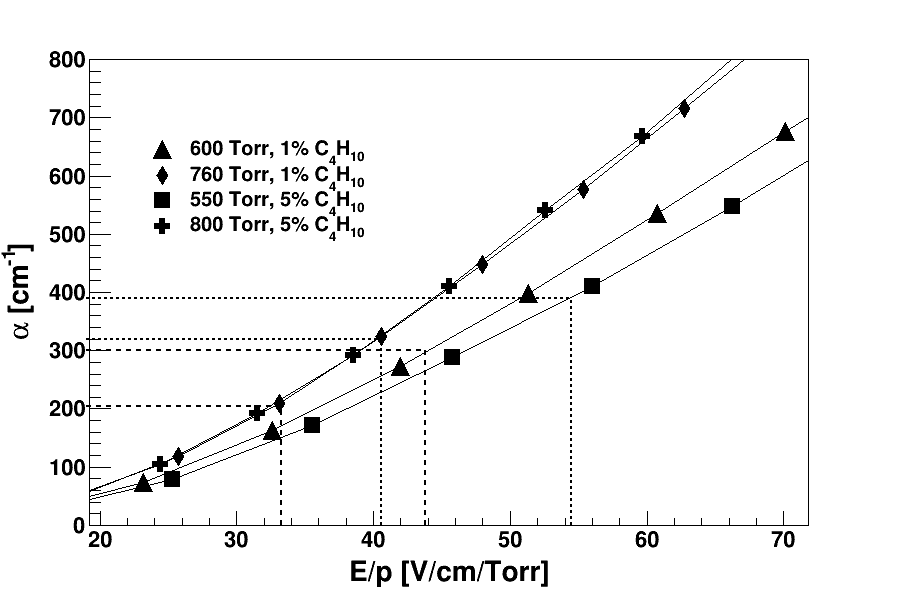}
\caption{\label{fig:gIso} A Magboltz simulation of the first Townsend coefficient, $\alpha$, as a function of the reduced field, $E/p$, for several Ar--C$_4$H$_{10}$ mixtures.  Higher gains can be achieved before sparking occurs in both 1\% and 5\% C$_4$H$_{10}$ mixtures. (dashed lines).}
\end{figure}

 A total of 6 configurations of C$_4$H$_{10}$/Ar gas mixtures were studied with different pressures, concentrations and gain (Table \ref{isobutane_table}).  The standard operating configuration for the fissionTPC when making cross section measurements is a mixture of 95\% Ar and 5\% C$_4$H$_{10}$ at $p=550$~Torr.  This mixture and pressure are stable in beam, optimally utilize the detector active area for fission fragment and $\alpha$-particle tracks, and provide high gain at a relatively low MICROMEGAS bias voltage.  A 5\% C$_4$H$_{10}$ mixture was also tested at $p=800$~Torr, and the MICROMEGAS was found to be unstable for $E/p > 39.8$.  A Magboltz simulation indicates that this onset of instability at $p=800$~Torr is at a lower gain than the configurations found to be stable at $p=550$~Torr (Fig.~\ref{fig:gIso}).  
 
 Additionally, a 1\% C$_4$H$_{10}$ mixture was tested at $p=600$~Torr for comparison with the data point reported in Table~\ref{results_table}.  The decreasing pressure was again found to allow for a higher gain (Fig.~\ref{fig:gIso}).  These finding  generally  agree with the CH$_4$ results.  Decreasing the charge density allows for a higher gain to be achieved before reaching the Raether limit.  It should be noted that Ar/C$_4$H$_{10}$ is a Penning mixture and its gain will be substantially higher than what is calculated directly using the $\alpha$ value~\cite{Sahin} from the Magboltz simulation. 

Stability against sparking was achieved with lower concentrations of C$_4$H$_{10}$ compared to CH$_4$, indicating that photon quenching plays a role since C$_4$H$_{10}$ has a higher photoabsorption cross section~\cite{Kameta}.

\begin{table}[htb]
\begin{center}
\begin{tabular}{cccc}
\hline
\rule{0pt}{3ex}%
\vspace{0.1cm}
\textbf{p (Torr)} & \textbf{Conc.} & $\mathbf{E/p}$ (V/cm/Torr) & \textbf{Stable}\\
\hline
\rule{0pt}{3ex}%
600 & 1\% & 43.8 & Yes\\
\rule{0pt}{3ex}%
760 & 1\% & 33.3 & Yes\\
\rule{0pt}{3ex}%
550 & 5\% & 48.5 & Yes \\
\rule{0pt}{3ex}%
550 & 5\% & 54.5 & Yes \\
800 & 5\% & 39.8 & Yes \\
\rule{0pt}{3ex}%
800 & 5\% & 40.6 & No \\
\end{tabular}
\end{center}
\caption{Ar--C$_4$H$_{10}$ mixtures and conditions tested for stability against sparking.}\label{isobutane_table}
\end{table}

\subsection{CO$_2$}
CO$_2$ is often used for its low diffusion and in systems requiring a non-flammable gas.  The Ar--CO$_2$ mixture was stable against sparking with lower concentrations and higher gain compared to the Ar--CH$_4$ mixture (Table \ref{co2_table}).  CO$_2$ has lower diffusion coefficients compared to CH$_4$, and is effective at cooling electrons~\cite{Bittl} which results in higher density charge clouds.  It also has a higher photoabsorption cross section~\cite{Kameta,Shaw} compared to CH$_4$.  In particular, its photoabsorption cross section is well matched to emission lines from excited states of Ar~\cite{Sahin2}.  The stability of the Ar-CO$_2$ mixture against sparking further indicates that the quenching of photons plays an important role in allowing in-beam operation.

\begin{table}[htb]
\begin{center}
\begin{tabular}{cccc}
\hline
\rule{0pt}{3ex}%
\vspace{0.1cm}
\textbf{p (Torr)} & \textbf{Conc.} & $\mathbf{E/p}$ (V/cm/Torr) & \textbf{Stable}\\
\hline
\rule{0pt}{3ex}%
600 & 1\% & 62.0 & Yes\\
\rule{0pt}{3ex}%
600 & 1\% & 71.8 & Yes\\
\rule{0pt}{3ex}%
760 & 1\% & 49.7 & Yes\\
\rule{0pt}{3ex}%
760 & 1\% & 50.6 & Yes\\
\rule{0pt}{3ex}%
900 & 1\% & 45.6 & Yes\\
\end{tabular}
\end{center}
\caption{Configurations and results for stability against sparking for an Ar--CO$_2$ mixture.}\label{co2_table}
\end{table}

\subsection{CF$_4$}
CF$_4$ is useful as a drift gas in TPCs and wire chambers because of its fast electron drift speed, which improves performance for high rate applications.  CF$_4$ was found to be a much less effective quench gas when compared to CO$_2$ and C$_4$H$_{10}$.  The 25\% CF$_4$ mixture was on the edge of stability in the these tests.  A slight increase in gain from the stable configuration at $p=760$~Torr shown in Table \ref{results_table} resulted in sparking (Table \ref{cf4_table}), which continued for a 30\% mixture at similar gain.  A decrease in pressure to $p=600$~Torr allows for a higher gain to be achieved, as was seen for CH$_4$ and C$_4$H$_{10}$.

In electron avalanches, CF$_4$ produces a large number of photons~\cite{Kaboth}, but it also has a low photoabsorption cross section to those photons~\cite{Pansky}.  While this is useful for scintillation, it limits the effectiveness of CF$_4$ as a quench gas.  The relative instability of the CF$_4$ gas mixtures when compared to C$_4$H$_{10}$ and CO$_2$ further supports the conclusion that photon quenching is an important factor in preventing sparking of the MICROMEGAS when in beam.

\begin{table}[htb]
\begin{center}
\begin{tabular}{cccc}
\hline
\rule{0pt}{3ex}%
\vspace{0.1cm}
\textbf{p (Torr)} & \textbf{Conc.} & $\mathbf{E/p}$ (V/cm/Torr) & \textbf{Stable}\\
\hline
\rule{0pt}{3ex}%
600 & 30\% & 75.2 & Yes\\
\rule{0pt}{3ex}%
760 & 25\% & 61.2 & Yes\\
\rule{0pt}{3ex}%
760 & 25\% & 62.6 & No\\
\rule{0pt}{3ex}%
760 & 30\% & 66.2 & No\\
\end{tabular}
\end{center}
\caption{Configurations and results for stability against sparking for an Ar--CF$_4$ mixture.}\label{cf4_table}
\end{table}

\section{Discussion} \label{discussion}
Sparking in a MICROMEGAS can result from spontaneous breakdown due to mesh defects, rate-induced effects, and highly ionizing particles~\cite{Bay}.  Defects that cause spontaneous breakdowns essentially limit the maximum voltage any particular MICROMEGAS can sustain.  In this study, each of the MICROMEGAS operating configurations was tested both in and out of beam to identify the  cause of sparking.  Rate-induced sparking is unlikely to be the cause since the beam-induced rate accounts for less than 2\% of the total charge generated in the gain stage. This was measured via the current on the mesh, which primarily results from the spontaneous $\alpha$-decay of the $^{239}$Pu source.  
Fission fragments are the highest energy events observed in the fissionTPC and produce an exceptionally large amount of total ionization.  They were eliminated as a cause by testing each of the gases (excepting CF$_4$) with a $^{252}$Cf spontaneous fission source in the fissionTPC.  It was observed that gases with good photon-quenching characteristics only required concentrations as low as 1\% to be stable in beam. Overall, reducing gas pressure also allowed the MICROMEGAS to achieve higher gains while maintaining stability.  

\begin{figure}[ht]
\centering\includegraphics[width=1.\linewidth]{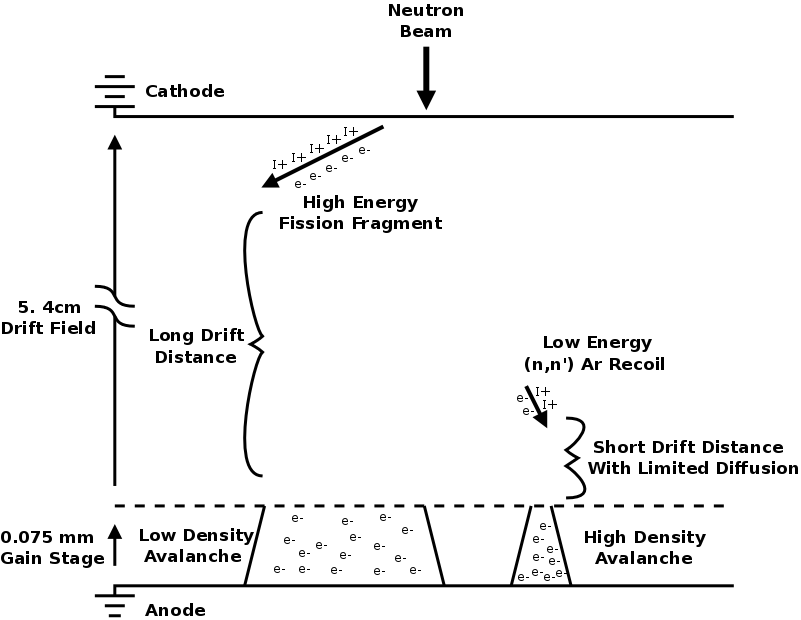}
\caption{\label{fig:schem} A schematic representation of the fissionTPC drift field and gain stage. The neutron beam passes directly through the TPC along the drift direction.  Fission fragments are only produced at the cathode and the resulting ionization drifts the full length, undergoing significant diffusion which reduces the ionization density entering the MICROMEGAS.  Conversely, nuclear recoils from neutron scattering can occur throughout the volume. When one of these interactions occurs near the mesh, the high-density ionization cloud enters the MICROMEGAS largely unaffected by diffusion.}
\end{figure}

\begin{figure}[ht]
\centering\includegraphics[width=1.\linewidth]{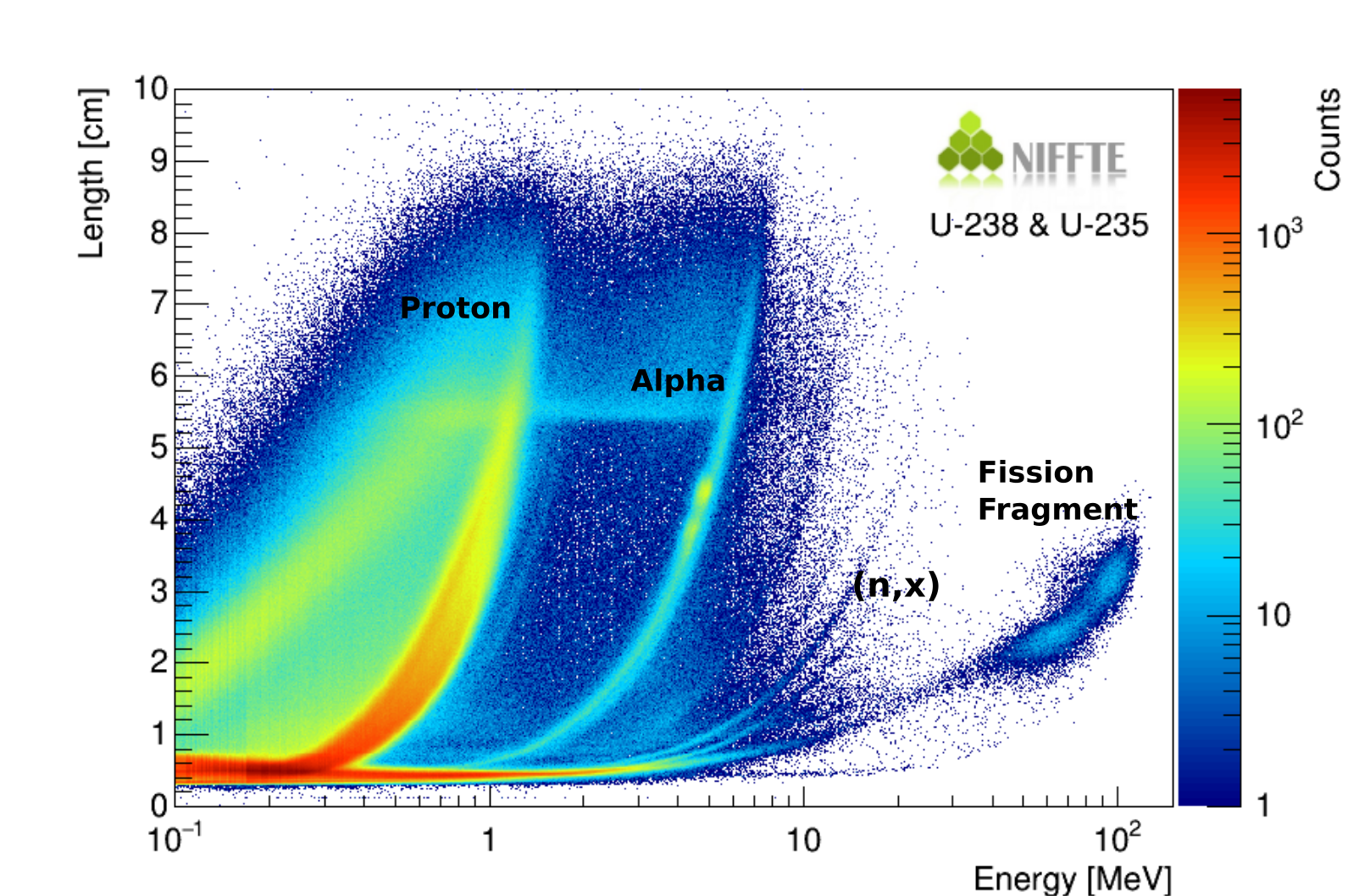}
\caption{\label{fig:gLvE} Track length versus track energy measured by the fissionTPC in-beam at LANSCE with an actinide source.  The gas was a mixture of 95\% Ar and 5\% C$_4$H$_{10}$ at 550 torr.  The proton, $\alpha$-particle, recoil and fission fragment bands are visible.  Events between the $\alpha$ and p bands are the result of missing channels and/or track reconstruction errors.  Events at energies lower than the proton band are due to partial reconstruction of proton tracks leaving the active area of the detector.}
\end{figure}

We conclude that the limiting factor for sparking in the MICROMEGAS is the density of the primary ionization clouds resulting from Ar recoils near the surface of the gain stage. A schematic representation of the fissionTPC and beam-induced events can be seen in Fig.~\ref{fig:schem}.  While recoils deposit a small amount of total ionization as compared to fission fragments, the ionization is deposited in a small volume of gas. Figure~\ref{fig:gLvE} shows a plot of track length versus energy for fissionTPC data collected with a mixture of 95\% Ar and 5\% C$_4$H$_{10}$.  Between the fragment and $\alpha$-particle bands are several bands due to recoils of other nuclei present in the fisstionTPC gas and detector material.  Between energies of 2--3 MeV, the bands become indistinguishable and drop below minimum track length resolution of the the fissionTPC which is limited by the 2~mm pixel size.  These recoil events can produce as many primary electrons as an $\alpha$ or proton track in a fraction of the length.  

If the diffusion time is limited, a relatively low-energy gas recoil can have a greater charge density than a fission fragment of much higher energy.  With a high enough space-charge density, a local electric field gradient can be produced in the gain stage that is sufficiently large to induce a discharge.  This is the same effect that is referred to as exceeding the Raether limit.  While the limit is commonly described in terms of a number of electrons, that is in reference to the gain when starting with only one electron.  The actual mechanism described by Raether~\cite{Raether} corresponds to  the space-charge of an avalanche being sufficient to develop a highly conductive plasma streamer directed at the anode. This is usually referred to as the \emph{rapid} or \emph{streamer} mechanism. Gas ionizing photons produced in the avalanche contribute to the portion of the streamer directed towards the cathode. This picture agrees with  the observed improvements in stability when operating with the more effective photon-quenching gases C$_4$H$_{10}$ and CO$_2$.   

The effect of diffusion on charge density was calculated using SRIM and Magboltz ~\cite{Biagi,Ziegler} for various energies and drift distances of two relevant cases: a typical fission fragment and Ar ions. A $100~$MeV $^{99}$Mo fission fragment in $1~$atm of argon gas has an ionization density at the start of the track (where ionization is highest for fission fragments and low energy heavy ions) more than 10 times that of a $1~$MeV Ar ion in the same gas.  The fission fragments in the fissionTPC originate only from the central cathode and have a drift distance of 5.4 cm for charge deposited at the start of the track, while Ar recoils can occur at any point along the drift axis.  Comparing the charge density within 1$\sigma$ of the transverse diffusion radius, assuming a uniform distribution, one finds that the charge produced by a $1~$MeV Ar ion drifting 1 mm has an ionization density at the beginning of the track approximately 5 times that of a fission fragment that has drifted 5.4 cm. Increasing the Ar ion energy to $10~$MeV increases the ionization density by a factor 20. This is supported by the observed improvement in stability  when operating at reduced pressures, which generally decreases the ionization density by extending the track length. 

In Section~\ref{p10section} it was reported that no discharges were observed when setting the drift field to $0~V/cm$ while maintaining an applied field across the gain stage.  If discharges are a result of Ar recoils, sparking should in principle still occur in this configuration.  
The stability criteria was defined for this experiment as 60 minutes of no observed discharges (see Section~\ref{procedures}). The $0~V/cm$ drift field data point meet this criteria.  For all the discharges that were recorded in this experiment the average time between being exposed to beam and observing a spark was 10 minutes, with no discharge taking more that 20 minutes to occur.  While the full observation period is 6 times longer that the average discharge rate, the 75um gain stage only represents 0.14\% of the total volume and correspondingly of the Ar recoil events.  It is therefore plausible that no $0~V/cm$ discharges occurred during the observation period.

Furthermore when the drift field is absent, only the fraction of charge deposited directly in the gap contributes to the avalanche, since all charge outside the gap would not drift.  For example a 1 MeV Ar recoil has a particle range of over 2.5mm in 760 torr of Argon gas, and deposits less than 10\% of its total charge in the first 100 um of track length.  While our argument is that charge density is the leading factor in inducing a spark, it is likely that discharges are not only a function of charge density but also of total charge in the event. 

\begin{figure}[ht]
\centering\includegraphics[width=1.\linewidth]{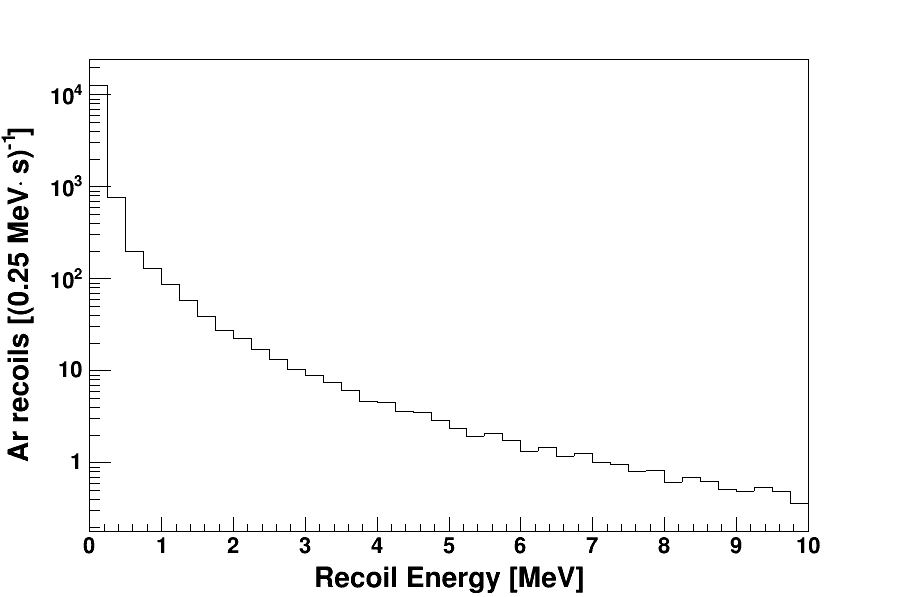}
\caption{\label{fig:ArRe} The expected rate of Ar recoils vs the recoil energy in the fissionTPC.  The figure was generated using a detailed MCNP simulation of the fissionTPC in the 90L beam line of WNR-LANSCE.  The simulated gas was the standard mixture for the fissionTPC of 95\% Ar and 5\% C$_4$H$_{10}$ at $550$~torr.}
\end{figure}

This experiment was a measure of a threshold effect whereby the MICROMEGAS either discharged or was stable for an extended time, as was described in section~\ref{procedures}.  The discharging rate varies strongly and non-linearly with gain.  Without a much more detailed simulation of the discharge process or dedicated data observations it is unclear which specific recoil energies over which drift distances lead to discharges for a given gas mixture and gain.  We did however run a simulation to provide a semi-quantitative check on whether Ar recoils were a plausible cause for discharging at the rates observed.  Figure~\ref{fig:ArRe} shows the expected rate and energy of Ar recoils in one volume of the fissionTPC with a gas mixture of 95\% Ar and 5\% C$_4$H$_{10}$ at a pressure of $550$~torr.  The results are from an MCNP simulation of the fissionTPC in the 90L beam line at WNR-LANSCE.  The total rate of Ar recoils over all energies is approximately $14000$~per~second.  As a specific example previously in this section it was discussed that a $1~$MeV Ar recoil drifting $1$~mm results in an ionization density 5 times that of a fission fragment. The rate of Ar recoils in the energy range of $0.75$--$1.25$~MeV within $1$~mm of the surface of the MICROMEGAS is approximately $4$~per~second.  The simulation shows that there are a sufficient number of high-energy Ar recoils expected in the fissionTPC to produce the observed discharges.

If Ar recoils are the primary cause of sparking, gas mixtures made of lighter noble gases such as helium or neon should prove to be more stable.  An initial observation indicates that the fissionTPC is able to achieve gains approximately twice as high in a Ne--C$_4$H$_{10}$ mixture, however a systematic study has not been conducted at this time. While a similar conclusion was reached in Ref.~\cite{Delbart}, a direct comparison with these measurements is difficult since the beam, target, pressures and gains used were in a considerably different regime than the fissionTPC operating conditions.

\section{Conclusions}
We have studied the stability of a MICROMEGAS against sparking when operated in a high-energy high-flux neutron beam.  Several Ar-based  gas mixtures relevant to TPC design were investigated.  Nuclear recoils from neutron scattering near the mesh, which result in high-density ionization tracks, were identified as  the cause of sparking instability.  When the recoil interactions occurs near the mesh, the diffusion distance is reduced and consequently the charge density of the ion cloud is high, leading to the Raether limit being exceeded. Quench gases C$_4$H$_{10}$ and CO$_2$ with high photoabsorption cross sections required lower concentrations to achieve stability against breakdown as compared to CH$_4$ and CF$_4$.  Overall, reducing the pressure of the operating gas allowed for higher operating gains. Use of gas mixtures based on nuclides lighter than Ar should also provide  stability at high gains for MICROMEGAS in neutron beams.  

\section{Acknowledgments}
The neutron beam for this work was provided by LANSCE, which is funded by the U.S. Department of Energy and operated by Los Alamos National Security, LLC, under contract DE-AC52-06NA25396.  This work performed under the auspices of the U.S Department of Energy by Lawrence Livermore National Laboratory under contract DE-AC52-07NA27344. This material is based upon work supported by the U.S. Department of Energy, National Nuclear Security Administration, Stewardship Science Academic Alliances Program, under Award Number DE-NA0002921. LLNL-JRNL-722401

\bibliographystyle{elsarticle-num}
\bibliography{references_Long_v2}

\end{document}